\documentclass{article}

     \usepackage[nonatbib,preprint]{mystyle}

\usepackage[utf8]{inputenc} 
\usepackage[T1]{fontenc}    
\usepackage{hyperref}       
\usepackage{amsfonts}       
\usepackage{microtype}      
\usepackage[cmex10]{amsmath}
\usepackage{amssymb,amsfonts,bbm,euscript}       
\usepackage{mathabx}
\usepackage{graphicx}

\newcommand{\Exp}{\mathsf{E}}

\newcommand{\f}{\mathsf{f}}
\newcommand{\g}{\mathsf{g}}
\newcommand{\h}{\mathsf{h}}
\newcommand{\p}{\mathsf{p}}
\newcommand{\G}{\mathsf{G}}
\newcommand{\T}{\mathsf{T}}
\newcommand{\C}{\mathsf{C}}

\newcommand{\cA}{\EuScript{A}}
\newcommand{\cB}{\EuScript{B}}
\newcommand{\cC}{\EuScript{C}}
\newcommand{\cX}{\EuScript{T}}
\newcommand{\cG}{\EuScript{G}}

\title{Image Restoration from Parametric Transformations using Generative Models}

\author{%
  Kalliopi~Basioti\\ 
  Department of Computer Science\\
  Rutgers University\\
  Piscataway, NJ 08854, USA \\
  \texttt{kib21@scarletmail.rutgers.edu} \\
   \And
   George V. Moustakides\\
  Department of Electrical and Computer Engineering\\
  University of Patras\\
  26500 Rion, Patras, Greece\\
  \texttt{moustaki@upatras.gr}\\
}

\begin{document}

\maketitle

\begin{abstract}
When images are statistically described by a generative model we can use this information to develop optimum techniques for various image restoration problems as inpainting, super-resolution, image coloring, generative model inversion, etc. With the help of the generative model it is possible to formulate, in a natural way, these restoration problems as Statistical estimation problems. Our approach, by combining maximum a-posteriori probability with maximum likelihood estimation, is capable of restoring images that are distorted by transformations even when the latter contain unknown parameters. The resulting optimization is completely defined with no parameters requiring tuning. This must be compared with the current state of the art which requires exact knowledge of the transformations and contains regularizer terms with weights that must be properly defined. Finally, we must mention that we extend our method to accommodate mixtures of multiple images where each image is described by its own generative model and we are able of successfully separating each participating image from a single mixture. 
\end{abstract}


\section{Introduction}
As a first step towards the presentation of our methodology for image restoration let us introduce a simple mathematical problem. Assume that we are observing a vector $Y$ which is a transformed and noisy version of a hidden vector $X$. We are interested in estimating $X$ from the observation $Y$ when we have available a generative model that captures the statistical behavior of $X$.

A number of well known computer vision problems can be formulated as the estimation problem we just described: \textit{Inpainting}, which consists in reconstructing an image $X$ when we have available its partially erased version $Y$; \textit{Super-resolution}, where from a lower resolution image $Y$ we recover a higher resolution version $X$; \textit{Image coloring}, where from a gray-color image $Y$ we recover the full-color (RGB) version $X$; \textit{Image separation} when we separate images from a linear (or nonlinear) combination and finally, \textit{Generative model inversion}, where from an image $Y$ we identify the input to the generative model that generates an output which is as close as possible to the available image $Y$; are just a few of the image restoration problems of interest.

Regarding image inpainting, classical techniques can be found in \cite{afonso, barnes, efros, hu, shen}; for super-resolution an overview of classical methods exists in \cite{allebach, freeman-1, freeman-2, gu, li-x} and for coloring in \cite{horiuchi, semary}. With the advent of generative networks \cite{goodfellow}, a new class of tools has emerged which is based on generative models. For inpainting such methods are considered in \cite{siavelis,yeh1,yeh2}, with the super-resolution and coloring problem also being mentioned in the same articles. The inversion of a generative model enjoys exactly the same formulation and suitable solutions can be found in \cite{daras,lei}. Finally, for the image separation problem recent efforts based on generative modeling can be found in \cite{anirudh,kong,soltani,subakan}.

Early efforts based on generative modeling \cite{daras,kong,lei,siavelis,yeh1,yeh2} were using the generative model \textit{partially} (only the generator function). Only recently \cite{anirudh,asim,soltani,subakan,whang} we see techniques that employ the \textit{complete} model (generator function and input density) improving the performance of the corresponding methods. Even in these recent efforts we observe the existence of weighting parameters that need to be tuned. The tuning process is carried out by applying the corresponding method \textit{multiple times} with different parameter values and adopting the one delivering the best result. Since our own methodology relies on the well established \textit{Statistical estimation theory}, we will be able to identify in a well defined way all parameters entering into the problem. In fact the method we are going to develop will be able to treat cases where the transformation responsible for the image distortion is \textit{not exactly known} as required by all existing methods.

\section{Statistical estimation theory}
Let us recall some very basic elements from Statistical estimation theory. Consider two random vectors $Z,Y$ where $Z$ is hidden while $Y$ is observed. Using $Y$, we would like to \textit{estimate} the hidden vector $Z$ when $Z,Y$ are statistically related, with their relationship captured by the joint probability density function (pdf) $\f(Y,Z)$. 

The existence of a joint density $\f(Y,Z)$ allows for the applications of the well known Bayesian estimation theory \cite[pp. 259–279]{veeravalli} for the estimation of $Z$ from $Y$. According to this theory, any deterministic function $\hat{Z}(Y)$ of $Y$ can play the role of an estimator for $Z$. Bayesian estimation provides a solid mathematical mechanism for evaluating these candidate estimators and identifying the one that will play the role of the final optimum estimator. Following this theory, we first need to provide a cost function $\C(\hat{Z},Z)$ that places a cost on each combination $\{\hat{Z},Z\}$ of estimate and true value. Then, the performance criterion of an estimator is defined as the average cost $\Exp[\C(\hat{Z}(Y),Z)]$, where expectation is with respect to $Z$ and $Y$. An estimator $\hat{Z}(Y)$ will be regarded as optimum if it minimizes the average cost among all deterministic functions of $Y$.

In the existing theory, one can find several meaningful cost functions along with their corresponding optimum estimators. From the available possibilities we distinguish two candidates that are of interest to our analysis. Specifically, we focus on the minimum mean square error (MMSE) and the maximum a-posteriori probability (MAP) estimators \cite[pp.~267-268]{veeravalli} which we present next.

\noindent\textbf{MMSE:} The cost function for this estimator is $\C(\hat{Z},Z)=\|\hat{Z}-Z\|^2$. It is then well known \cite{veeravalli} that the optimum estimator is defined as the conditional mean
\begin{equation}
\hat{Z}=\Exp[Z|Y]=\int Z \f(Z|Y)\,dZ=\int Z \frac{\f(Y,Z)}{\f(Y)}\,dZ=\frac{\int Z\f(Y,Z)\,dZ}{\int \f(Y,Z)\,dZ}.
\label{eq:1.1}
\end{equation}
where $\f(Z|Y)$ denotes the conditional probability density of $Z$ given $Y$.

\noindent\textbf{MAP:} Here the cost function is somewhat peculiar and defined as
$$
\C(\hat{Z},Z)=\left\{\begin{array}{cl}
1,&\text{if}~\|\hat{Z}-Z\|\geq\delta\\
0,&\text{otherwise},
\end{array}\right.
$$
where $0<\delta\ll1$ denotes a very small quantity (tending to 0). This criterion is known \cite[page 267]{veeravalli} to lead to the well known MAP estimator which is defined as
\begin{equation}
\hat{Z}=\text{arg}\max_Z\f(Z|Y)=\text{arg}\max_Z\frac{\f(Y,Z)}{\f(Y)}=\text{arg}\max_Z\f(Y,Z),
\label{eq:1.2}
\end{equation}
corresponding to the \textit{most likely} $Z$ given the observations $Y$. There are of course other popular alternatives as, for example, the minimum mean absolute error (MMAE) criterion which leads to the conditional median estimator \cite[page 267]{veeravalli}. However, in this work we analyze only the two estimators depicted in \eqref{eq:1.1} and \eqref{eq:1.2} and in the simulations we basically use the MAP estimator that presents clear computational advantages.

\subsection{Including unknown parameters}
The previous classical results are based on the assumption that there is available (known) a joint density $\f(Y,Z)$ that captures the statistical relationship between $Z$ and $Y$. In practice, however, the joint density may also contain a number of parameters that we express with the help of a vector $\gamma$. In other words, the joint pdf of $Y,Z$ is of the form $\f(Y,Z|\gamma)$ for some vector $\gamma$. The Bayesian approach treats parameters as random as well, consequently $\f(Y,Z|\gamma)$ is regarded as the pdf of $Y,Z$ \textit{conditioned} on $\gamma$. Since $\gamma$ is also random, its statistical behavior is expressed by a pdf $\p(\gamma)$. This implies that the joint density of all three random vectors has the form $\f(Y,Z,\gamma)=\f(Y,Z|\gamma)\p(\gamma)$. 

As before, we assume that we observe $Y$ and interested in estimating $Z$. The question now is how should we treat $\gamma$. There are two possibilities:

\underline{\textit{Marginalization of $\gamma$}:} We can compute $\f(Y,Z)=\int\f(Y,Z|\gamma)\p(\gamma)\,d\gamma$ and then use \eqref{eq:1.1} or \eqref{eq:1.2}.

\underline{\textit{Estimation of $\gamma$:}} We can apply \eqref{eq:1.1} or \eqref{eq:1.2} with $\hat{Z}$ replaced by $\{\hat{Z},\hat{\gamma}\}$, thus considering $\gamma$ as part of the quantities to be estimated. Since our focus is only on $\hat{Z}$ this implies that finding $\hat{\gamma}$ is simply an intermediate step towards our desired goal.

In the applications of interest, as we will see, we have available the density $\f(Y,Z|\gamma)$ but \textit{not} $\p(\gamma)$. We can overcome this lack of knowledge by following a \textit{worst-case scenario}, namely assume that $\p(\gamma)$ is the \textit{most uninformative prior}. If $\gamma\in\cC$, where $\cC$ is some known set, then this corresponds to selecting $\p(\gamma)$ to be the \textit{uniform} over $\cC$, provided that the Lebesgue measure $\mu(\cC)$ is finite. If $\mu(\cC)=\infty$ then we can adopt for $\p(\gamma)$ the \textit{improper uniform} \cite[Page 27]{oconnor}. 
We can easily verify that in both cases we obtain the same results if we consider from the start that $\p(\gamma)=1$ for all $\gamma\in\cC$. This is exactly what we are going to use in our subsequent analysis.

\noindent\textbf{MMSE:} Here, marginalization and estimation of $\gamma$ result in exactly the same estimate for $Z$ which, under the improper uniform, takes the form
\begin{equation}
\bar{\f}(Y,Z)=\int\limits_{\gamma\in\cC}\f(Y,Z|\gamma)\,d\gamma,~~
\hat{Z}=\frac{\int Z\bar{\f}(Y,Z)\,dZ}{\int\bar{\f}(Y,Z)\,dZ}. 
\label{eq:1.4-2}
\end{equation}
\textbf{MAP:} For this estimator the two approaches differ. In the case of the marginalization approach we obtain
\begin{equation}
\hat{Z}=\text{arg}\max_{Z}\bar{\f}(Z,Y),
\label{eq:1.5-1}
\end{equation}
with $\bar{\f}(Z,Y)$ defined in \eqref{eq:1.4-2}, while the estimation approach yields
\begin{equation}
\tilde{\f}(Z,Y)=\max_{\gamma\in\cC}\f(Y,Z|\gamma),~~\hat{Z}=\text{arg}\max_{Z}\tilde{\f}(Z,Y).
\label{eq:1.5-2}
\end{equation}
We observe that \eqref{eq:1.5-2} is equivalent to first performing a \textit{maximum likelihood} estimation \cite[pp. 319–351]{veeravalli} of $\gamma$ followed by a MAP estimation of $Z$. 

After this very brief presentation of the necessary results from Statistical estimation theory, we are now ready to apply these ideas to image restoration problems.

\section{Image restoration and generative model}
Let us focus on the general problem of interest and include the generative model into our formulation. Suppose $X$ is a random vector described by the generative model $X=\G(Z)$ with the input $Z$ being distributed according to the density $\h(Z)$. Both functions $\{\G(Z),\h(Z)\}$ that comprise the generative model are assumed \textit{known}. The generator $\G(Z)$ can be a neural network (deep or shallow), trained with the help of adversarial \cite{arjovsky,basioti1,binkowski,goodfellow,nowozin} or non-adversarial techniques \cite{basioti2,dziugaite}. If a non-adversarial training method is adopted then this clearly suggests that a discriminator function, which plays an important role in the techniques proposed in \cite{daras,siavelis,yeh1,yeh2}, does not exist. For this reason our goal, similarly to \cite{asim,bora,whang}, is to propose estimation techniques that do not rely on discriminator functions. 

Every time a realization of $X$ occurs, we assume that it undergoes a transformation and its noisy version is observed as a data vector $Y$. More specifically, the observation vector $Y$ and the original vector $X$ are related through the following equation
\begin{equation}
Y=\T(X,\alpha)+W.
\label{eq:2.1}
\end{equation}
$\T(X,\alpha)$ is a deterministic transformation with known functional form that can possibly contain \textit{unknown} parameters $\alpha\in\cA$ with $\cA$ a known set.\footnote{When parameter vector $\alpha$ exists we assume that it may change with \textit{each realization} of $X$. For this reason $\alpha$ \textit{cannot} be tuned before hand using training data and make $\T(X,\alpha)$ completely known.}
$W$ is a random vector independent from $X$ that expresses additive noise and/or \textit{modeling errors}. For $W$ we assume that it is distributed according to the density $\g(W|\beta)$ which has a known functional form and possibly \textit{unknown} parameters $\beta\in\cB$, with $\cB$ a known set. The problem we would like to solve is the recovery of the vector $X$ from the observation vector $Y$. To achieve this estimate we intend to exploit the generative model in the following way: Instead of finding directly the estimate $\hat{X}$ as in \cite{asim,whang}, 
we propose to obtain the estimate $\hat{Z}$ of the input to the generator and then estimate $X$ as $\hat{X}=\G(\hat{Z})$. This simple variation allows for the adoption of \textit{any form} for the generator function without special preference to invertible ones and, as we will see, for efficient handling of unknown parameters.

To apply the estimation theory presented in the previous section we first need to find the joint density of $Y,Z$. It is easy to see that the parameter vector $\gamma$ of the general theory corresponds to the combination $\{\alpha,\beta\}$ and
\begin{equation}
\allowdisplaybreaks 
\f(Y,Z|\alpha,\beta)=\g\big(Y-\T(\G(Z),\alpha)|\beta\big)\,\h(Z).
\label{eq:2.3}
\end{equation}
In \eqref{eq:2.3} $\h(Z)$ is completely known since it is the pdf of the input to the generative model, while $\g(W|\beta)$ is known up to possibly some unknown parameter vector $\beta$. If we apply the estimators of the previous section then for the MMSE in \eqref{eq:1.4-2} we must define
\begin{equation}
\bar{\g}(W)=\int\limits_{\beta\in\cB}\g(W|\beta)\,d\beta,~~~~
\bar{\bar{\g}}(Y,Z)=\int\limits_{\alpha\in\cA}\bar{\g}\big(Y-\T(\G(Z),\alpha)\big)\,d\alpha
\label{eq:2.4.5}
\end{equation}
which yields the estimate
\begin{equation}
\hat{Z}=\frac{\int Z\bar{\bar{\g}}(Y,Z)\h(Z)\,dZ}{\int \bar{\bar{\g}}(Y,Z)\h(Z)\,dZ}
=\frac{\Exp_{Z}[Z\bar{\bar{\g}}(Y,Z)]}{\Exp_{Z}[\bar{\bar{\g}}(Y,Z)]}.
\label{eq:2.4}
\end{equation}
The last ratio contains expectations with respect to $Z$ which is distributed according to $\h(Z)$. Similarly, for the MAP estimator in \eqref{eq:1.5-1} and \eqref{eq:1.5-2} we can write for the marginalization approach that
\begin{equation}
\hat{Z}=\text{arg}\max_Z\bar{\bar{\g}}(Y,Z),
\label{eq:2.5-1}
\end{equation}
with $\bar{\bar{\g}}(Y,Z)$ defined in \eqref{eq:2.4.5}, while for the estimation approach we have
\begin{equation}
\tilde{\g}(W)=\max_{\beta\in\cB} \g(W|\beta),~~
\tilde{\tilde{\g}}(Y,Z)=\max_{\alpha\in\cA} \tilde{\g}\big(Y-\T(\G(Z),\alpha)\big),~~
\hat{Z}=\text{arg}\max_Z\tilde{\tilde{\g}}(Y,Z)\h(Z).
\label{eq:2.5-2}
\end{equation}
When the transformation $\T(X)$ does not contain any unknown parameters, the previous expressions simplify to
\begin{equation}
\hat{Z}=\frac{\int Z\bar{\g}\big(Y-\T(\G(Z))\big)\h(Z)\,dZ}{\int \bar{\g}\big(Y-\T(\G(Z))\big)\h(Z)\,dZ}
=\frac{\Exp_{Z}\big[Z\bar{\g}\big(Y-\T(\G(Z))\big]}{\Exp_{Z}\big[\bar{\g}\big(Y-\T(\G(Z))\big)\big]},
\label{eq:2.4.55}
\end{equation}
for the MMSE, while for the MAP estimation we have
\begin{equation}
\hat{Z}=\text{arg}\max_Z\bar{\g}\big(Y-\T(\G(Z))\big)\h(Z),~\text{or}~
\hat{Z}=\text{arg}\max_Z\tilde{\g}\big(Y-\T(\G(Z))\big)\h(Z),\label{eq:2.5-10}
\end{equation}
with the first corresponding to marginalization and the second to maximum likelihood estimation of $\beta$.
Our approach, because it is based on the classical Statistical estimation theory, enjoys a number of interesting properties: 1)~As in \cite{asim,whang}, it uses the \textit{complete} generator model in order to perform the estimation and does not require any discriminator function. 2)~The final optimization problem does not contain terms in the form of regularizers that include \textit{unknown weights} that require tuning. 3)~We can treat transformations and noise pdfs that contain unknown parameters which are being properly identified using maximum likelihood estimation. The last two properties are unique to our proposed approach and are not present in any other existing generative image restoration methodology.

\subsection{Gaussian noise and Gaussian input}
Let us now specify in more detail our mathematical model. For the additive noise vector $W$ appearing in the data model in \eqref{eq:2.1}, we assume that it has Gaussian elements that are independent and identically distributed with mean zero and variance $\beta^2$. We adopt the Gaussian model only for simplicity. It is possible to resort to more general noise models as for example the Student's $t$-distribution which was successfully employed in classical (non generative) image restoration techniques \cite{chantas}. Unfortunately Student's $t$ distribution does not offer closed form expressions for its parameters as the Gaussian case. And this is something we would like to have in order to be able to compare our resulting cost function with the costs employed in the existing literature. Limiting ourselves to Gaussian noise for $W$ where each element has zero mean and variance $\beta^2$, yields
$$
\g(W|\beta)\!=\!\frac{e^{-\frac{1}{2\beta^2}\|W\|^2}}{\sqrt{(2\pi)^N\beta^{2N}}},~~
\bar{\g}(W)\!=\!\int_0^\infty\g(W|\beta)\,d\beta\!=\!\frac{C}{\|W\|^{N+1}},~~\tilde{\g}(W)\!=\!\max_{\beta\geq0}\g(W|\beta)\!=\!\frac{C'}{\|W\|^N},
$$
where $C,C'$ constants and $N$ is the length of the vector $W$. If we assume that the input density $\h(Z)$ is also Gaussian $\mathcal{N}(0,I)$, where $I$ the identity matrix, then for known transformations $\T(X)$ the MMSE estimate in \eqref{eq:2.4.5} becomes
\begin{equation}
\hat{Z}=\frac{\Exp_{Z}[Z\|Y-\T(\G(Z))\|^{-(N+1)}]}{\Exp_{Z}[\|Y-\T(\G(Z))\|^{-(N+1)}]}\approx
\frac{\sum_{i=1}^LZ_i\|Y-\T(\G(Z_i))\|^{-(N+1)}]}{\sum_{i=1}^L\|Y-\T(\G(Z_i))\|^{-(N+1)}]}.
\label{eq:mbofla}
\end{equation}
We note that we generate realizations $\{Z_1,\ldots,Z_L\}$ of $\h(Z)$ and by evoking the the Law of Large Numbers we approximate the MMSE estimate. For the MAP estimates in \eqref{eq:2.5-10} we have
\begin{equation}
\hat{Z}=\text{arg}\min_Z\{M\log\|Y-\T(\G(Z))\|^2+\|Z\|^2\},~\text{where}~M=N+1~\text{or}~M=N,
\label{eq:mbifla}
\end{equation}
with $M=N+1$ corresponding to the marginalization and $M=N$ to the estimation of $\beta$. The two expressions are clearly very similar especially in the case where $N\gg1$. We can now compare our optimizations in \eqref{eq:mbifla} with 
\begin{equation}
\hat{Z}=\text{arg}\min_Z\{\|Y-\T(\G(Z))\|^2+\lambda\|Z\|\}~\text{and}~\hat{Z}=\text{arg}\min_Z\{\|Y-\T(\G(Z))\|^2+\lambda\|Z\|^2\},
\label{eq:mbifla-2}
\end{equation}
where the first is proposed in \cite{asim} and the second in \cite{bora,whang}. Both approaches in \eqref{eq:mbifla-2} contain an \textit{unknown weight} $\lambda$ which in order to be tuned properly we need to solve the corresponding optimization problem several times for different values of this parameter and select the one delivering the best results. In our approach in \eqref{eq:mbifla} such parameter is clearly \textit{not present}. Another notable difference is how the error distance $\|Y-\T(\G(Z))\|^2$ is combined with the input power $\|Z\|^2$. In our method we use the \textit{logarithm} of the distance while in \cite{asim,bora,whang} it is the distance itself combined with $\|Z\|$ or $\|Z\|^2$. We would like to emphasize that the cost function of our approach is not selected in some arbitrary way but it is the outcome of a theoretical analysis which is based on the Statistical estimation theory.

Even though the MMSE estimator in \eqref{eq:mbofla} does not involve any additional optimization when the transformation $\T(X)$ has no parameters, it can be used only when the length $N$ of $W$ is small. Indeed, for large $N$ the expression $\|Y-\T(\G(Z_i))\|^{-(N+1)}$ may very easily take extremely small or extremely large values which will cause computational problems due to finite precision. Unless $N$ is of the order of a few tens this method should be avoided. This observation clearly applies to images where $N$ can be several thousands. From now on we focus on MAP estimation and since the difference between the two versions in \eqref{eq:mbifla} is minor when $N$ is large, we adopt the second version where we estimate $\beta$ instead of marginalizing it.

\subsection{Parametric transformations}
Let us now consider the more challenging problem of a transformation $\T(X,\alpha)$ containing unknown parameters $\alpha$. Following our general theory developed for the case of noise and generator input being Gaussian, the MAP estimator with maximum likelihood estimation of the parameters is equivalent to
\begin{equation}
\hat{Z}=\text{arg}\min_Z\big\{N\log\big(\min_{\alpha}\|Y-\T(\G(Z),\alpha)\|^2\big)+\|Z\|^2\big\}.
\label{eq:mbofla1}
\end{equation}
Additionally, if the transformation is linear\footnote{In most restoration problems the transformation enters as a matrix multiplied element-by-element with the ideal image which is also expressed as a matrix. If we reshape the image into a vector then this multiplication becomes a classical linear matrix/vector multiplication which is what we adopt in our analysis.}, that is, $\T(X,\alpha)=\T(\alpha)X$, which is the case in most restoration problems, then the optimization problem in \eqref{eq:mbofla1} becomes
\begin{equation}
\hat{Z}=\text{arg}\min_Z\big\{N\log\big(\min_{\alpha}\|Y-\T(\alpha)\G(Z)\|^2\big)+\|Z\|^2\big\},
\label{eq:mbofla2}
\end{equation}
where $\T(\alpha)$ is a matrix parametrized with $\alpha$.
Finally we can further advance our analysis if we assume that $\T(\alpha)$ is linear in its parameters namely it can be decomposed as 
\begin{equation}
\T(\alpha)=\alpha_1\T_1+\cdots+\alpha_M\T_M
\label{eq:mbofla3-00}
\end{equation}
where $\T_1,\ldots,T_M$ are known matrices and $\alpha=[\alpha_1,\ldots,\alpha_M]^{\intercal}$ is the unknown parameter vector. As an example consider the coloring problem where we have $\T_{\rm R},\T_{\rm G},\T_{\rm B}$ with each matrix isolating the corresponding RGB component from the ideal colored image and the scalar quantities $\{\alpha_{\rm R},\alpha_{\rm G},\alpha_{\rm B}\}$ denoting the percentage by which each component contributes to the final gray level. If these percentages are known before hand then the resulting transformation $\T=\alpha_{\rm R}\T_{\rm R}+\alpha_{\rm G}\T_{\rm G}+\alpha_{\rm B}\T_{\rm B}$ is also known and the coloring problem can be treated by existing techniques. If, however, $\{\alpha_{\rm R},\alpha_{\rm G},\alpha_{\rm B}\}$ are \textit{unknown}, then we need to estimate these parameters in parallel with $Z$, by solving \eqref{eq:mbofla2}.

Regarding the minimization with respect to $\alpha$ we can either combine it with the minimization with respect to $Z$ and use, for example, a gradient descent for the pair $\{Z,\alpha\}$ or, in the linear case we can find the analytic solution of the minimization with respect to $\alpha$, substitute it, and then minimize only over $Z$. The first idea is straightforward and requires no further explanation. For the second if we define the matrix $\cX=[\T_1\G(Z),\ldots,\T_M\G(Z)]$ then from \eqref{eq:mbofla2} and \eqref{eq:mbofla3-00} we conclude that
$$
\min_{\alpha}\|Y-\T(\alpha)\G(Z)\|^2=\min_{\alpha}\|Y-\cX\alpha\|^2=\|Y\|^2-Y^{\intercal}\cX(\cX^{\intercal}\cX)^{-1}\cX^{\intercal}Y,
$$
with the last outcome following from the Orthogonality Principle \cite{hajek} and expressing the projection error of $Y$ onto the space generated by the columns of $\cX$. This result, when substituted in \eqref{eq:mbofla2}, yields
\begin{equation}
\hat{Z}=\text{arg}\min_Z\left\{N\log\left(\|Y\|^2-Y^{\intercal}\cX(\cX^{\intercal}\cX)^{-1}\cX^{\intercal}Y\right)+\|Z\|^2\right\},
\label{eq:mbofla3}
\end{equation}
where the only unknown is $Z$ and, we recall from its definition, that $\cX$ is also a function of $Z$.

\subsection{The image separation problem}
In \cite{anirudh,soltani,subakan} existing single image methods are extended to combinations of \textit{multiple} images where each image is described by a separate generative model. These extensions experience the same drawbacks as the original methods, namely, 1)~they contain multiple regularizer terms with unknown weights that need to be properly determined, 2)~the proposed criteria are not the outcome of some rigorous mathematical analysis, and 3)~the corresponding methods in \cite{anirudh,soltani,subakan} cannot accommodate combinations involving unknown parameters. 

We can overcome the previous weaknesses by generalizing our methodology to cover multiple images as well. For simplicity we only consider the two image case with the analysis of any number of images being very similar. Suppose that we have two images $X_1,X_2$ each satisfying a generative model $X_i=\G_i(Z_i)$ with input density $Z_i\sim\h_i(Z),~i=1,2$. If the data model follows
\begin{equation}
Y=\alpha_1X_1+\alpha_2X_2+W
\label{eq:model2}
\end{equation}
where the additive noise/modeling error $W$ has density $\g(W|\beta)$ with parameters $\beta$ then we can combine all parts and produce the joint probability density
\begin{equation}
\f(Y,Z_1,Z_2|\alpha_1,\alpha_2,\beta)=\g\big(Y-\alpha_1\G_1(Z_1)-\alpha_2\G_2(Z_2)|\beta\big)\,\h_1(Z_1)\,\h_2(Z_2).
\label{eq:mbaraskouak}
\end{equation}
In \eqref{eq:mbaraskouak} we made the assumption that $Z_1,Z_2$ are statistically independent which produced the product of the two input densities. Following the general theory and limiting ourselves to the MAP estimator we need to solve the optimization problem
\begin{equation}
\max_{Z_1,Z_2}\max_{\alpha_1,\alpha_2,\beta}\!\f(Y,Z_1,Z_2|\alpha_1,\alpha_2,\beta)
\!=\!\max_{Z_1,Z_2}\!\big\{\!\max_{\alpha_1,\alpha_2,\beta}\g\big(Y-\alpha_1\G_1(Z_1)-\alpha_2\G_2(Z_2)|\beta\big)\big\}\h_1(Z_1)\h_2(Z_2).
\label{eq:mbic1}
\end{equation}
If, as before, $\g(W|\beta)$ is Gaussian with mean 0 and covariance $\beta^2I$ and both input vectors are independent Gaussian with mean 0 and unit covariance matrix then from \eqref{eq:mbic1}, after maximizing over $\beta$, we conclude that
\begin{equation}
\{\hat{Z}_1,\hat{Z}_2\}=
\text{arg}\,\min_{Z_1,Z_2}\big\{N\log\big(\min_{\alpha_1,\alpha_2}\|Y-\alpha_1\G_1(Z_1)-\alpha_2\G_2(Z_2)\|^2\big)+\|Z_1\|^2+\|Z_2\|^2\big\}.
\label{eq:mbic2}
\end{equation}
We can either apply gradient descent on the combination $\{Z_1,Z_2,\alpha_1,\alpha_2\}$ or solve analytically for $\{\alpha_1,\alpha_2\}$, substitute, and then minimize over $\{Z_1,Z_2\}$. Regarding the latter we have from the Orthogonality Principle \cite{hajek}
$$
\min_{\alpha_1,\alpha_2}\|Y-\alpha_1\G_1(Z_1)-\alpha_2\G_2(Z_2)\|^2=\|Y\|^2-Y^{\intercal}\cG(\cG^{\intercal}\cG)^{-1}\cG^{\intercal}Y
$$
where $\cG=[\G_1(Z_1), \G_2(Z_2)]$. Substituting in \eqref{eq:mbic2} gives rise to
\begin{equation}
\{\hat{Z}_1,\hat{Z}_2\}=
\text{arg}\,\min_{Z_1,Z_2}\left\{N\log\left(\|Y\|^2-Y^{\intercal}\cG(\cG^{\intercal}\cG)^{-1}\cG^{\intercal}Y\right)+\|Z_1\|^2+\|Z_2\|^2\right\},
\label{eq:mbic3}
\end{equation}
and where we observe that $\cG$ is a function of $\{Z_1,Z_2\}$. 

We can also accommodate the case where transformations are applied to each individual image suggesting that each image can undergo a different deformation before the final mixture. This corresponds to replacing each component $\alpha_i\G_i(Z_i)$ in \eqref{eq:model2} with $\T_i(\G_i(Z_i),\alpha_i)$ where each transformation $\T_i(\cdot)$ can have its own unknown local parameters $\alpha_i$. Obtaining the necessary equations for this more general setup presents no particular difficulty.

Unlike the classical source separation problem \cite{cardoso,kofidis} where we need as many (linear) combinations of the sources as the number of sources we are seeking, here separation can be achieved from a \textit{single} mixture. Of course this is possible because we have available a \textit{statistical description} of the sources in the form of generative models. We recall that in classical source separation such description is not present and separation is achieved by processing directly the available combinations.

\section{Experiments}

\subsection{Datasets and Pretrained GAN models}\label{ssec:A1}
For our experiments, we use the CelebA \cite{liu} and the Caltech Birds \cite{wah} datasets. The first dataset contains 202,599 RGB images that are cropped and resized to 64x64x3 and then separated into two sets 202,499 for training and 100 for testing. For the Birds dataset, we train two models, one with the original images and the second with segmented images, namely, with removed background using the included segmentation masks. In both cases, the images are resized to 64x64x3 while we kept 10,609 images for training and 1179 for testing.

The training data are used to design Wasserstein GANs \cite{arjovsky} with the progressive method developed in \cite{karras}. For all cases we use the following configuration, Generator: input 512 Gaussians $\mathcal{N}(0,1)$ and five layers. Each layer consists of two convolutions with two kernels $3\times3$ except the first layer that has one $4\times4$ and one $3\times3$ kernel and the last that has two $3\times3$ and one $1\times1$ kernel resulting in an output of $64\times64\times3$. After each convolution, a leaky ReLU is applied except for the last $1\times1$ convolution where no nonlinear function is used. The intermediate layers also involve an upsampling operation. Discriminator: Input $64\times64\times3$ with six layers in total. The first five layers have two convolutions with two $3\times3$ kernels except for the first layer which has an additional $1\times1$ layer and the last layer which has a $3\times3$ and a $4\times4$ kernel. After each convolution, we apply a leaky ReLU except for the last $4\times4$ kernel where no nonlinearity is used. In the intermediate layers, we apply downsampling except for the last layer. Finally, we employ a fully connected part that provides the scalar output of the discriminator.

In all competing methods, we apply the momentum gradient descent \cite{qian} with normalized gradients where the momentum hyperparameter is set to 0.999, the learning rate to 0.001 and the algorithm runs for 200,000 iterations.

\subsection{Deblurring}
Perhaps the most common deformation is due to a linear filter convolved with an image. In particular when the filter is one-dimensional and applied to the image row-by-row then this can model a \textit{horizontal motion blur}. This idea can be clearly extended to cover blurring in any direction but for simplicity we consider the case of horizontal blurring. We use a finite impulse response filter of length 5 with coefficients $\alpha_1=1.0187;\alpha_2=-0.5933;\alpha_3=-0.3501;\alpha_4=0.4635;\alpha_5=-0.24$ that were randomly generated. The goal is from the deformed image to restore the original. We compare the methods of \cite{yeh1} and \cite{whang} (which in this example coincides with \cite{bora}) against our method. 

The techniques in \cite{yeh1,whang} \textit{require exact knowledge} of the filter coefficients. They also require fine-tuning of the weight $\lambda$ appearing in \eqref{eq:mbifla-2}. This is achieved by solving multiple instances of optimization problems with various weight values and selecting the one delivering the smallest reconstruction error. In the case of \cite{yeh1} this turns out to be 0.6 while in \cite{whang} 0.2. We should emphasize that these values are \textit{filter, transformation and data dependent} meaning that if the filter coefficients or the transformation or the class of data changes we need to repeat the tuning procedure. \textit{What is even more crucial is the fact that tuning requires exact knowledge of the filter. Consequently, if the filter contains unknown parameters, the tuning process is impossible.}

Since our method has no unknown weights it can be applied directly without the need of any fine-tuning phase. We distinguish two versions in our approach. In the first we assume that we know the filter coefficients exactly in order to make fair comparisons with \cite{yeh1,whang}. In the second version we assume that the filter coefficients are \textit{unknown} which implies that we simultaneously estimate the filter coefficients and restore the original image by solving \ref{eq:mbofla3}. Unlike our proposed technique, existing methods do not perform this combined optimization and are therefore \textit{unable to restore the original image when the transformation contains unknown parameters}.

\begin{figure}[h]
\centering
\includegraphics[width=\hsize]{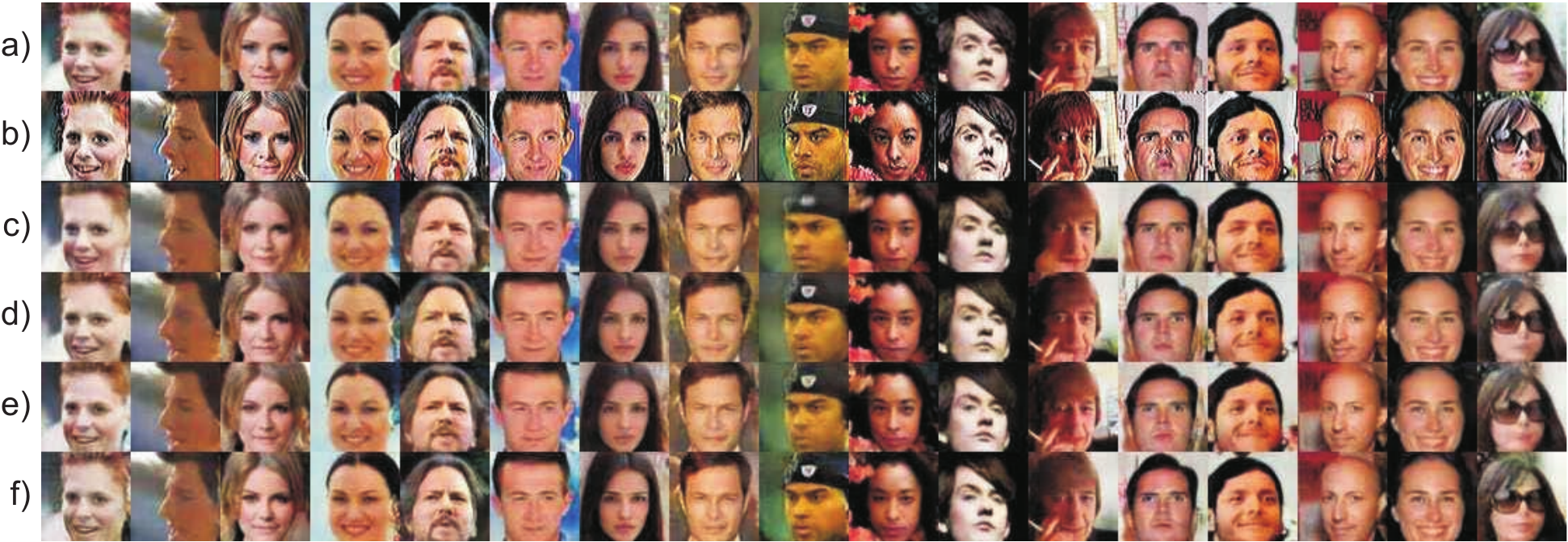}
\caption{Deblurring. Row a)~Original; b)~Blurred; c)~\cite{yeh1}, known parameters; d)~\cite{whang}, known parameters; e)~Proposed, known parameters; f)~Proposed, unknown parameters.}
\label{fig:1}
\end{figure}
\vskip0.2cm
\begin{table}[h]
\centering\begin{tabular}{|l|c|c|}
\hline
    Method    & Error   & PSNR      \\ \hline
\cite{yeh1}, known    & 0.0029 & 25.4742 \\ 
\cite{whang}, known  & 0.0021 & 26.7480 \\ 
Proposed, known   & 0.0022 & 26.4880 \\ 
Proposed, unknown & 0.0030 & 25.4944 \\ \hline
\end{tabular}
\vskip0.2cm
\caption{Reconstruction errors and PSNRs for Deblurring.}
\label{table:1}
\end{table}
As in \cite{daras, yeh1} we ran our simulations for every testing image three times, with different initializations and retain the solution with the smallest reconstruction error. In Figure\,\ref{fig:1} we present the corresponding results for the problem of horizontal blurring. Row a)~depicts the original faces; in row b) we see their blurred version when we apply the selected filter; rows c), d), e) present the restoration provided by \cite{yeh1}, \cite{whang} and our first version respectively when the filter is known; finally row f)~is the restoration results of our second version with the filter coefficients being unknown and estimated \textit{for each image} at the same time with the restoration process. Table\,\ref{table:1} contains the corresponding restoration errors\footnote{Smaller error does not necessarily imply that the method provides visually better results. If this were the case then the preferable optimization would have been to minimize the restoration error. However it is well known that this criterion does not lead to satisfactory restorations (see for example video \cite[Time 17:02]{nvidia}).} (per pixel average squared error between original and restored) and the Peak Signal to Noise Ratio (PSNR) of each method. 

Visibly it is difficult to distinguish differences between the various techniques. This fact is also captured in Table \ref{table:1} where the restoration errors and the PSNRs are comparable. The same observation applies even in the case of our second version that estimates the unknown coefficients. We must however emphasize once more that our versions compared to the existing techniques enjoy certain unique properties: 1)~There are no weights that need fine-tuning; 2)~The restoration quality is comparable even when the transformations have unknown parameters and 3)~Our criterion is not an ad-hoc selection but the outcome of a rigorous mathematical analysis.

\subsection{Colorization}
\begin{figure}[h] 
\centering
\includegraphics[width=\hsize]{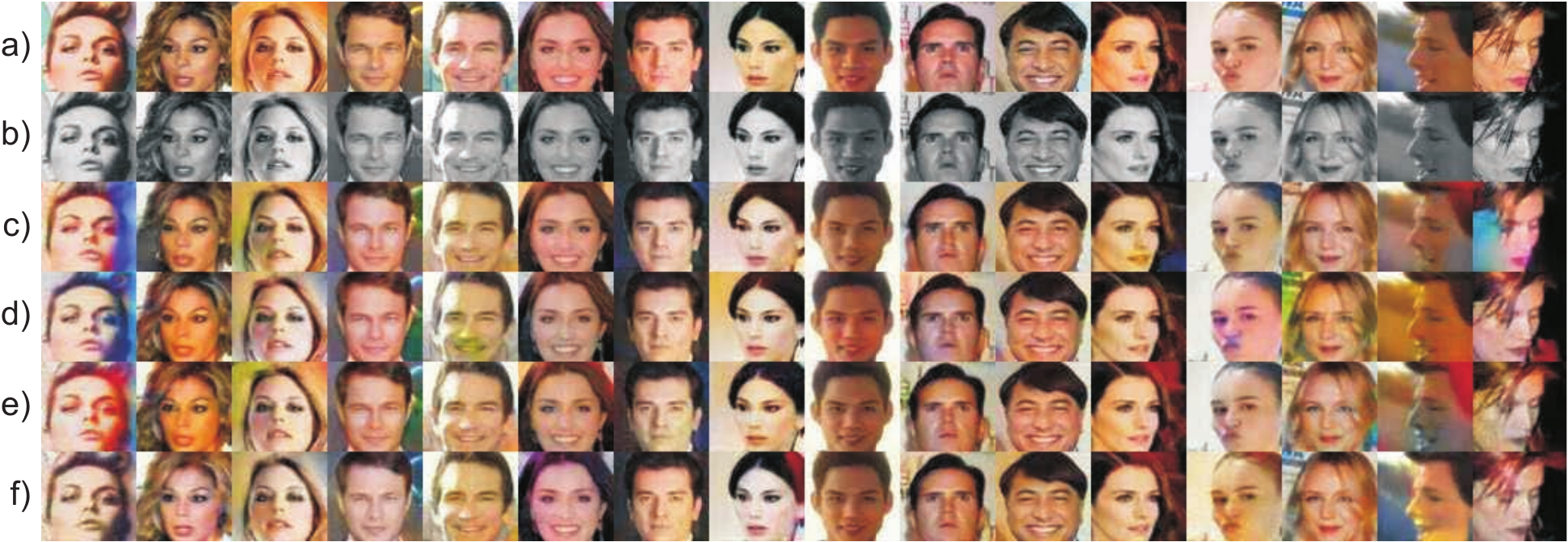}
\caption{Row a)~Original; b)~Transformed; c)~\cite{yeh1}, known parameters; d)~\cite{whang}, known parameters; e)~Proposed, known parameters; f)~Proposed, unknown parameters.}
\label{fig:2}
\end{figure}

In the second set of experiments we recover an RGB image from one of its chromatic components. As such we select the green channel. This information is passed onto the methods of \cite{yeh1,whang} and to our first version. In our second version we assume that we do not know which channel generates the observed gray-level data and we attempt to \textit{find the right channel} at the same time with the restoration process. We recall that the channel decomposition is a linear transformation that can be implemented with three matrices $\T_{\rm R},\T_{\rm G},\T_{\rm B}$. The fact that the unknown parameter is now \textit{discrete} does not pose any special difficulty in the optimization in \eqref{eq:mbofla2}, which must be modified as follows
\begin{equation}
\hat{Z}=\text{arg}\min_Z\big\{N\log\big(\min_{i={\rm R,G,B}}\|Y-\T_i\G(Z)\|^2\big)+\|Z\|^2\big\},
\label{eq:mbofla2newA}
\end{equation}
For the solution of \eqref{eq:mbifla}, \eqref{eq:mbifla-2} and \eqref{eq:mbofla2newA} we employ the same algorithm and hyperparameters as in our first set of experiments, except of course the weight $\lambda$ in \cite{yeh1} and \cite{whang} which we have to retune. This results in 0.1 for \cite{yeh1} and 0.5 for \cite{whang}.
In Figure \ref{fig:2}, row a) depicts the original RGB images; we see the green channel in gray-level in row b); rows c), d), e) have the restoration results of \cite{yeh1}, \cite{whang} and our proposed first version respectively
\vskip0.2cm
\begin{table}[h]
\centering
\begin{tabular}{|l|c|c|}
\hline
    Method    & Error   & PSNR      \\ \hline
\cite{yeh1}, known    & 0.0083                                                         & 21.5753 \\
\cite{whang}, known  & 0.0096                                                         & 21.3204 \\ 
Proposed, known   & 0.0089                                                         & 21.6724 \\ 
Proposed, unknown & 0.0093                                                         & 21.1516 \\ \hline
\end{tabular}
\vskip0.2cm
\caption{Reconstruction errors and PSNRs for Colorization.}
\label{tab:2}
\end{table}
when the channel is known; finally row f)~contains the results of our second version when the channel is unknown and must be discovered by solving \eqref{eq:mbofla2newA}. We also see in Table \ref{tab:2} the corresponding restoration errors and PSNRs. Again we realize that our proposed methodology provides comparable restoration quality as the existing methods when the transformation is known without the need to fine-tune any weights. Additionally, it also delivers similar quality even if the transformation contains unknown parameters which may take discrete or continuous values.

\subsection{Image Separation}
For the image separation problem we start by forming mixtures that are combinations of images from CelebA and segmented Caltech Birds. As before we implement the momentum gradient descent \cite{qian} with normalized gradients. The momentum hyperparameter is set to 0.999, the learning rate to 0.1 and we run the algorithm for 200,000 iterations.

When the mixture coefficients $\alpha_1, \alpha_2$ are known we compare our method with the method developed in \cite{soltani} which consists in solving the following optimization problem:
\begin{equation}
\{\hat{Z}_1, \hat{Z}_2\} = \text{arg}\min_{Z_1,Z_2}\big\{\|Y-\alpha_1G_1(Z_1) - \alpha_2 G_2(Z_2) \|^2+\lambda_1\|Z_1\|^2+\lambda_2\|Z_2\|^2\big\}.
\label{eq:mbofla2new}
\end{equation}
Weights $\lambda_1, \lambda_2$, as in the single image case, need to be fine-tuned. As before we select the values offering the best performance in terms of reconstruction error and PSNR. 

\begin{figure}[h]
\centering
\includegraphics[width=\hsize]{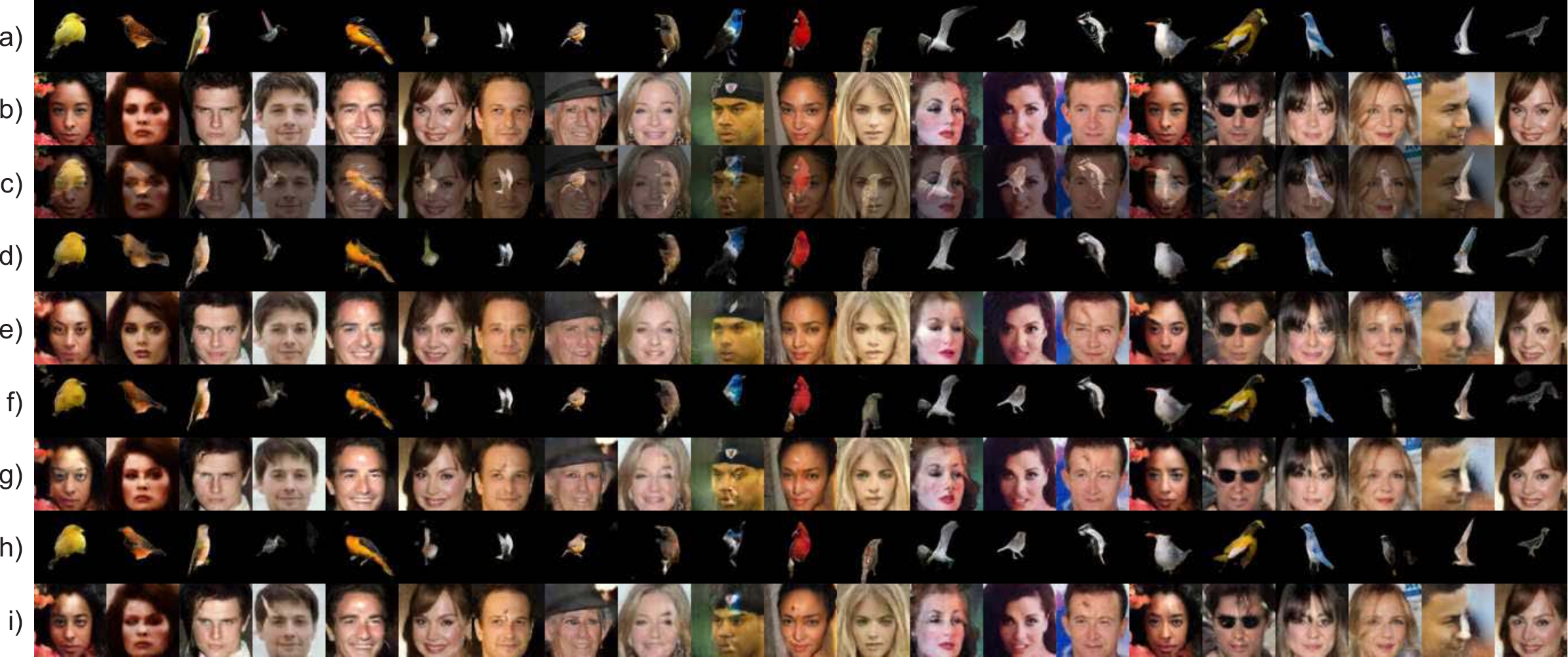}
\caption{\small Rows a,b)~Originals; c)~Mixture; d),e) \cite{soltani}, known coefficients; f),g) ~Proposed, known coefficients; h),i)~Proposed, unknown coefficients.}
\label{fig:3}
\end{figure}

For the case where $\alpha_1,\alpha_2$ are unknown, we select values satisfying the constraint $\alpha_1+\alpha_2=1$. Even though this is not necessary, we pass this information to our second version that estimates the two parameters in order to observe how it performs when there are also constraints present. Specifically we solve the problem
\begin{align*}
&\{\hat{Z}_1,\hat{Z}_2\}=
\text{arg}\,\min_{Z_1,Z_2}\big\{N\log\big(\min_{\alpha_1,\alpha_2}\|Y-\alpha_1\G_1(Z_1)-\alpha_2\G_2(Z_2)\|^2\big)+\|Z_1\|^2+\|Z_2\|^2\big\}\\ 
&\text{for\ } \alpha_1+\alpha_2=1.
\label{eq:mbic2}
\end{align*}
In the first set of experiments we select $\alpha_1=\alpha_2=0.5$. For \cite{soltani}, fine-tuning its parameters results in $\lambda_1=\lambda_2=0.3$. Figure\,\ref{fig:3} contains separation examples using the segmented version of the Caltech Birds. Specifically in rows a), b) we have the original images, in row c) their mixture, in d), e) the separated images by the method in \cite{soltani} when the mixture coefficients are known, in f), g) the corresponding results of our first version with the mixture coefficients being known and in h), i) the reconstructed images by our second version when the mixture coefficients are unknown. Table\,\ref{tab:3} depicts the corresponding reconstruction errors and PSNRs per dataset.

\begin{table}[h]
\centering
\begin{tabular}{|l|c|c|c|c|}
\hline
    Method    & Error (1) & Error (2)   & PSNR (1)  &  PSNR (2)   \\ \hline
\cite{soltani}, known & 0.0039 & 0.0065 & 23.1497 & 22.5328 \\ 
Proposed, known   & 0.0032 & 0.0039  & 25.2947 & 25.0748\\ 
Proposed, unknown & 0.0033 & 0.0048  & 25.1798& 24.6500\\ \hline
\end{tabular}
\vskip0.2cm
\caption{Reconstruction errors and PSNRs for the segmented Caltech birds (1) and CelebA (2) datasets when $\alpha_1 = \alpha_2 = 0.5$.}
\label{tab:3}
\vskip0.2cm
\centering
\begin{tabular}{|l|c|c|c|c|}
\hline
    Method    & Error (1) & Error (2)   & PSNR (1)  &  PSNR (2)   \\ \hline
\cite{soltani}, known & 0.0034 & 0.0051 & 23.9315 & 23.1957 \\ 
Proposed, known   & 0.0024 & 0.0054  & 26.6518 & 23.7610\\ 
Proposed, unknown & 0.0032 & 0.0079  & 26.0759 & 22.8339\\ \hline
\end{tabular}
\vskip0.2cm
\caption{Reconstruction errors and PSNRs for the segmented Caltech birds (1) and CelebA (2) datasets when $\alpha_1 = 0.6, \alpha_2 = 0.4$.}
\label{tab:4}
\end{table}
We also experimented with unequal mixing parameters and used $\alpha_1=0.6,\alpha_2=0.4$. 
As we mentioned before, every time the transformation changes the weights 
$\lambda_1,\lambda_2$ defining the optimization problem in \eqref{eq:mbofla2new}, for the method in \cite{soltani}, need to be retuned. This time tuning came up with the values $\lambda_1=\lambda_2 = 0.1$. In Table \ref{tab:4} we show the reconstruction errors and PSNRs achieved by \cite{soltani} and our first version where $\alpha_1,\alpha_2$ are known and also our second version that treats $\alpha_1,\alpha_2$ as unknown.

Finally, in our last set of experiments we separate CelebA 
faces from the original Caltech Birds. We use again $\alpha_1=\alpha_2=0.5$. Since the datasets have changed, the weights $\lambda_1, \lambda_2$ in \cite{soltani} need to be retuned. This time the best values we obtain are $\lambda_1=0.5, \lambda_2=0.4$. Due to the more complicated nature of the images in the two datasets, for convergence we increased the number of iterations to 400,000. 

\begin{figure}[h]
\centering
\includegraphics[width=\hsize]{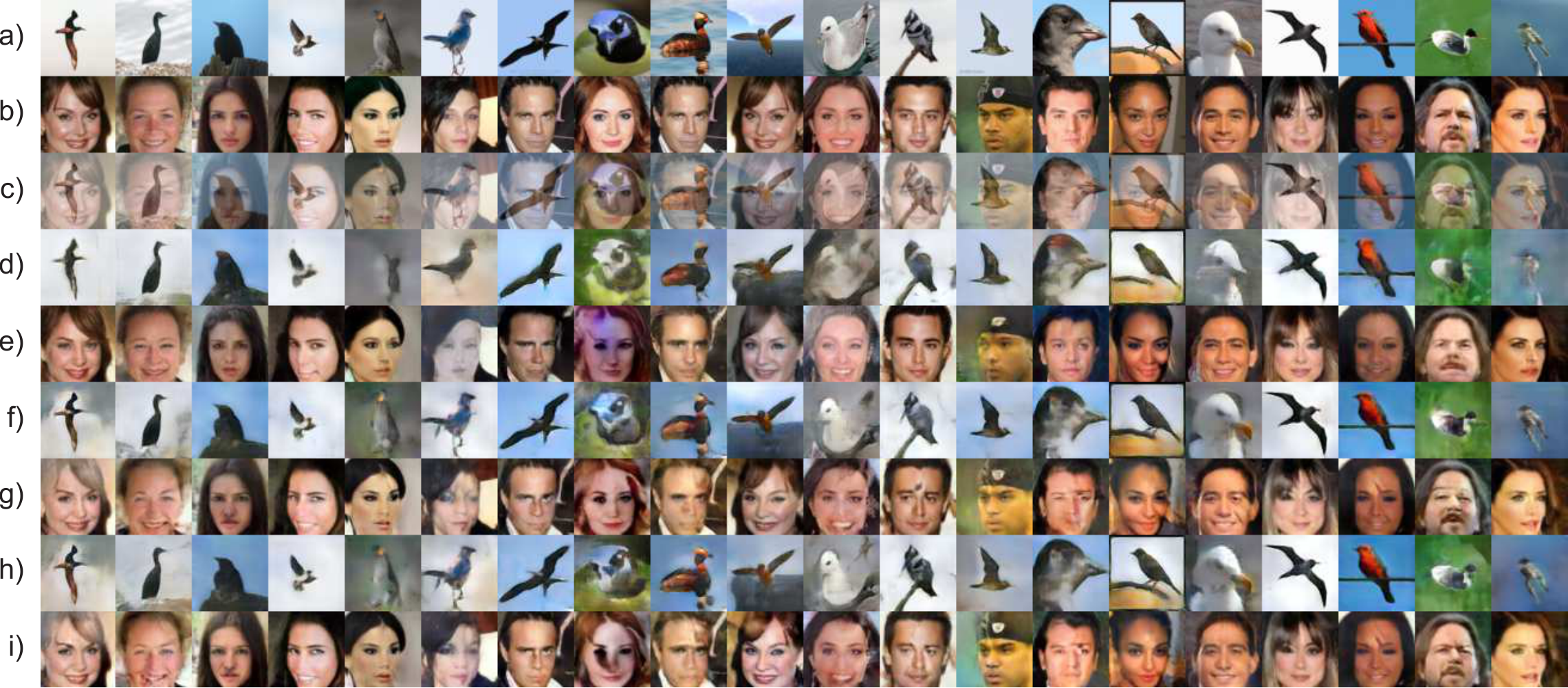}
\caption{Rows a,b)~Originals; c)~Mixture; d),e) \cite{soltani}, known coefficients; f),g) ~Proposed, known coefficients; h),i)~Proposed, unknown coefficients.}
\label{fig:4}
\end{figure}
\vskip0.2cm
\begin{table}[h!]
\centering
\begin{tabular}{|l|c|c|c|c|}
\hline
    Method    & Error (1) & Error (2)   & PSNR (1)  &  PSNR (2)   \\ \hline
\cite{soltani}, known & 0.0179 & 0.0181 & 18.2011 & 18.4105 \\
Proposed, known   & 0.0178 & 0.0180  & 18.4584 & 19.0702\\ 
Proposed, unknown & 0.0228 & 0.0281  & 17.7269 & 18.2589\\ \hline
\end{tabular}
\vskip0.2cm
\caption{Reconstruction errors and PSNRs for the Caltech birds (1) and CelebA (2) datasets when $\alpha_1 = \alpha_2 = 0.5$.}
\label{tab:5}
\vskip-0.5cm
\end{table}
In Figure\,\ref{fig:4} the rows a), b) have the original images, row c) their mixture, rows d) ,e) the separated images by \cite{soltani} for known mixture coefficients, rows f), g) depict the results of our first version where mixture coefficients are known and rows h), i) the corresponding reconstruction by our second version when the coefficients are unknown. In Table \ref{tab:5} we can see the corresponding reconstruction errors and PSNRs per dataset. 

\section{Conclusion}
We introduced a general image restoration methodology which is based on generative model description of the class (or classes) of images to be restored. Our processing technique relies on the Statistical estimation theory and is capable of restoring images through a mathematically well-defined optimization problem that does not require any tuning of weights of regularizer terms. The most important advantage of our method consists in its ability to restore and/or separate images even when the transformations responsible for the deformation contain unknown parameters. Experiments using popular dataset show that our technique, when applied to transformations with unknown parameters, it is capable of delivering similar restoration quality as the existing state of the art that needs exact knowledge of the transformations and tuning of weights for regularizer terms.

\subsubsection*{Acknowledgment}
\vskip-0.2cm 
This work was supported by the US National Science Foundation under Grant CIF\,1513373, through Rutgers University.

\end{document}